\documentclass[aps,showpacs,superscriptadress,preprint]{revtex4}
\usepackage{graphicx}
\usepackage{amsmath}
\begin{document}

\centerline{\bf An approach to chemical freeze-out scenario of identified particle spectra}
\centerline{\bf at 200AGeV Au-Au collisions at RHIC}

\vspace{12pt}

\centerline{ { Wei-Liang Qian}$^{\rm a}$, { Rone Andrade}$^{\rm a}$, {Fr\'{e}d\'{e}rique Grassi}$^{\rm a}$, {Yogiro Hama}$^{\rm a}$ and {Takeshi Kodama}$^{\rm b}$}

\vspace{12pt}

\centerline{$^{\rm a}$Instituto de F\'{\i}sica, USP} 
\centerline{C. P. 66318, 05315-970 S\~{a}o Paulo, SP, Brazil}

\vspace{12pt}

\centerline{$^{\rm b}$Instituto de F\'{\i}sica, UFRJ} 
\centerline{C. P. 68528, 21945-970 Rio de Janeiro-RJ , Brazil} 

\vspace{12pt} \vspace{12pt}

\begin{abstract}
Thermal model fit indicates early chemical freeze-out of multi-strange hadrons
with small collective velocities at 200AGeV Au-Au collisions at RHIC.
In this work, we present our recent results by SPheRIO hydrodynamical calculations 
inspired by this picture.
In our model, multi-strange hadrons go through chemical
freeze-out when the system reaches some temperature close to the phase transition,
stopping to make inelastic collisions, and their abundances are 
therefore determined only by partonic EOS. 
At a lower temperature thermal freeze-out takes place
where elastic collisions are brought to a halt.
We calculate the spectra for various hadrons at different centrality windows, 
with chemical and thermal freeze-out temperature being fit as a function of centrality.
As it is shown, the result provides a reasonable panoramic description of the
spectra of identified particles. 
Chemical freeze-out gives good correction of the multiplicity of certain species of particles, 
especially for multi-strange hadrons.
\end{abstract}
\vspace{12pt}

\maketitle

\section{Introduction}
In heavy-ion collision a hot and dense matter 
is formed which eventually evolves into a state of freely 
streaming particles. 
Strong collective flow patterns 
measured at RHIC suggest that 
the hydrodynamical model is well justified during the 
intermediate stages of the reaction. 
The model provides a useful tool for 
drawing information of space-time evolution of the matter, 
from the time when local equilibrium is attained until 
the process of hadron decoupling, called freeze-out.
The simplest scenario to treat freeze-out is to adopt chemical/thermal freeze-out temperature\cite{tc1,tc2},
it is widely used to give qualitative estimation of overall properties of system. 
However, it is worth noting that conclusions drawn about characteristic of chemical/thermal freeze-out temperature 
are essentialy based on statistics and hydrodynamics-motivated model fittings.
(i) chemical freeze-out: 
Statistics and hydrodynamics-motivated models give us
the characteristic temperature of $T_{ch}$ in relativistic heavy-ion collisions.
It is determined from fits of particle yield ratios for hadrons to the 
experimental data\cite{ch1,ch2,ch3,ch4,ch5,ch6,ch7,ch8,ch9,ch10,ch11,ch12,ch13,ch14,ch15}.
The temperature indicates when the hadron abundances become fixed. 
Fits are done for various centrality windows, and reveal that $T_{ch}$ is almost independent of centrality\cite{ch9,ch11,ch12,ch13,ch14,ch15}.
For 200AGeV Au-Au collisions, the value of chemical freeze-out temperature is around 160MeV, close to the phase boundary.
However, it is usually assumed strangeness is not fully equilibrated, 
thus a strangeness saturation factor ${\gamma}_{S}$ is introduced. 
Unlike chemical freeze-out temperature, ${\gamma}_{S}$ varies with collision system and centrality\cite{ch13,ch15}.
The compilation of chemical freeze-out parameters ($T_{ch} ,{\mu}^{B}_{ch}$) 
at various collision energies seem to be aligned on one curve in the $T-{\mu}^{B}$ plane. 
And can be parametrized by the average energy per particle $<E>/<N>{\sim}1$ GeV\cite{ch4}.
(ii) thermal (kinetic) freeze-out: 
The thermal (kinetic) freeze-out temperature $T_{th}$ 
is obtained from the slope of transverse momentum distribution with some 
radial flow profile. 
On the thermal freeze-out surface, elastic rescattering processes 
cease and hadrons start to escape freely. 
Statistical model fit shows that temperature at the kinetic freeze-out depends 
on the centrality\cite{ch13,ch15,th1,th2,th3,th4}. 
It is generally understood that the system undergoes first the chemical freeze-out where the observed 
particle ratios are fixed and next the thermal freezeout 
where the shape of the transverse-momentum distribution is fixed. 
Thus $T_{ch}{\geq}T_{th}$ is expected. 

It is worthwhile to check to what extent chemical/thermal freeze-out picture as well as above conclusions can be confirmed 
by the more realistic and accurate full hydrodynamic model, 
where the space-time evolution of fluid is taken into account rigorously. 
So in this work we explore the scenario of chemical freeze-out by calculating
transverse momentum spectra of 200A GeV Au-Au collisions. 
A full three-dimensional hydrodynamical model with chemical freeze-out mechanism is employed.
The results are compared with experimental data for all centrality windows.

The paper is organized as follows. In the next section, we briefly review the 
hydrodynamic model we used and explain how chemical freeze-out is implemented.
Our results and discussions are presented in the last section.

\section{SPheRIO with chemical freeze-out}

The hydrodynamical model we adopted is based on smoothed particle hydrodynamic(SPH)
algorithm\cite{topics,va}, 
In this model, the matter flow is parametrized in terms of discrete Lagrangian coordinates, 
of the so-called SPH particles. As a result, the hydrodynamic equations are reduced to 
a system of coupled ordinary differential equations. 
The main advantage of the model is that it can 
tackle the problem with highly asymmetrical configurations, 
as is concerned in relativistic high-energy nucleus-nucleus collisions. 

The code which implements the entropy representation of the SPH model for
relativistic high energy collisions, and which has been developed within the 
S\~{a}o Paulo - Rio de Janeiro Collaboration, is called SPheRIO. 
It has been successfully used to investigate the effects of the initial-condition fluctuations and adopting
the continuous emission scenario for the description of decoupling process 
\cite{va,topics,ce,ebe,v2}.

Usually, highly symmetric and smooth initial conditions is employed 
in hydrodynamic approach. 
However, owing to the fact that in heavy ion collisions the system is small, 
fluctuations are not negligible. 
To take this into account, 
we use an event simulator, NEXUS, to generate the initial condition of SPheRIO. 
This provides us possibility to study the collision on fluctuating event-by-event basis.

In the previous applications of SPheRIO, the strangeness had been neglected. 
Here, SPheRIO has been further improved in order to consider
strangeness conservation and to adopt the scenario of chemical freeze-out.
The strangeness conservation is implemented by explicitly incorporating 
strangeness chemical potential into the code, and correspondingly a new set of
equation of state(EOS) has been built and utilized. 
We use the hadronic resonance model with finite 
volume correction to describe the matter on the hadronic side, 
where the main part of observed resonances in Particle Data Tables\cite{pdtable} has been included. 
For quark gluon plasma (QGP) phase, the ideal gas model is adopted. 
As a good approximation, we assume local strangeness neutrality throughout the hydro evolution.

As it was shown\cite{qian}, hydrodynamic model without chemical freeze-out gives good
description of transverse momentum spectra of light hadrons.
While it reproduces the shape of spectra for hyperons and anti-proton, 
there are visible discrepancies in the multiplicities. 
To compensate this, we implement chemical freeze-out 
for strange hadrons such as $\Lambda$, $\Xi$, $\Omega$ and $\phi$ and anti-protons.
At chemical freeze-out temperature $T_{ch}$, 
these particles cease to have inelastic collisions and therefore their abundances are fixed.
We assume complete chemical equilibrium on the surface of chemical freeze-out, 
thus no strangeness saturation factor ${\gamma}_{S}$ is introduced in our treatment. 
With the system being cooled and rarefied further, 
thermal freeze-out occurs at a lower temperature $T_{th}$, 
where the distribution function and abundances of the rest of hadrons are determined.
In our calculation, $T_{ch}$ and $T_{th}$ serve as adjustable parameters
as functions of centrality according to experimental data.

\section{Results and discussions}

The thermal freeze-out temperature is tuned as a function
of centrality, to reproduce the shape of transverse momentum distribution of all charged particles.
Furthermore, as in the previous works \cite{va,topics,ce,ebe}, 
a rescaling factor is introduced to fix 
the pseudo-rapidity distribution for all charged particles.

In Fig.1, we computed the
pseudo-rapidity distribution for all charged particles for Au+Au collisions
at 200A GeV. It is depicted for six different centrality windows. 
The experimental data are from PHOBOS Collaboration\cite{phobos1}.

The experimental transverse-momentum-distribution data for all charged
particles is presented in Fig.2, the experimental data are reproduced with
a choice of freeze-out temperature from $T_{th} = 135$ MeV for most central collisions
to $T_{th} = 150$ MeV for most peripheral ones.
We use linear interpolation to determine the thermal freeze-out temperatures of 
intermediate windows.
The experimental data are from STAR Collaboration\cite{star1},
taken in Au+Au at 200A GeV, with $\eta=0$.

With the parameters chosen as explainded above, we calculate in the following 
the spectra for various hadrons.
In Fig.3-8 we show the transverse-momentum spectra 
of pions, protons and kaons for different centrality windows at mid rapidity, as well as
experimental data from BRAHMS Collaboration\cite{brahms1}.
The spectra of $\Lambda$, $\Xi$ and $\Omega$ are depicted in Fig.9-13,
together with data from STAR Collaboration\cite{star2}. We use dotted lines to represent
the results obtained without incorporating chemical freeze-out, and solid lines for
those with chemical freeze-out switched on. 
The interpolated values of thermal and chemical freeze-out temperatures are noted in the 
figures.

It is observed that the present hydrodynamic model gives good
description of the experimental transverse-momentum spectra for
pions, kaons and protons even without turning the chemical freeze-out on for these particles. 
Without the chemical freeze-out incorporated, although it gives the correct slopes 
for the spectra of anti-protons and strange hyperons $\Lambda$, $\Xi$ and $\Omega$,
the disagreement comes from the multiplicities of the spectra.
By introducing chemical freeze-out, those results are improved significantly.
While the spectra for pions, kaons and protons almost remain the same, (this is not shown in the figures.)
it provides a good fit of anti-proton and strange hyperons $\Lambda$, $\Xi$ and $\Omega$.
The chemical freeze-out increases the multiplicities of strange hadrons with respect to the ones in the original model 
owing to higher chemical freeze-out temperature, as compared with the thermal freeze out one.
Meanwhile, the slopes of the spectra remain unchanged.
It was indicated experimentally in Ref.\cite{kdis},
as one goes to large rapidity region where the baryon density differs sizably from zero, the outcome of chemical freeze-out would be much
more significant. Further work on this topic is under progress.

It is worth noting that the effects of chemical freeze-out have been discussed by several authors.
In ref\cite{hch1,hch2,hch3,hch4}, the early CFO is studied in terms of hybrid model
where the hydodynamic evolution is complemented with a hadronic cascade
model where the value of chemical freeze-out temperature is not studied explicitly. 
Other hydrodynamic calculations have been carried out either only
in the transverse plane assuming Bjorken's scaling\cite{hch5} or with the baryon chemical potential 
taken to be zero\cite{hch6}, and only some of the centrality windows are covered.

We acknowledge financial support by FAPESP (2004/10619-9, 2005/54595-9, CAPES/PrOBRAL, CNPq, FAPERJ and PRONEX.

\begin{figure}[!htb]
\vspace*{-1cm}
\includegraphics[width=8.5cm]{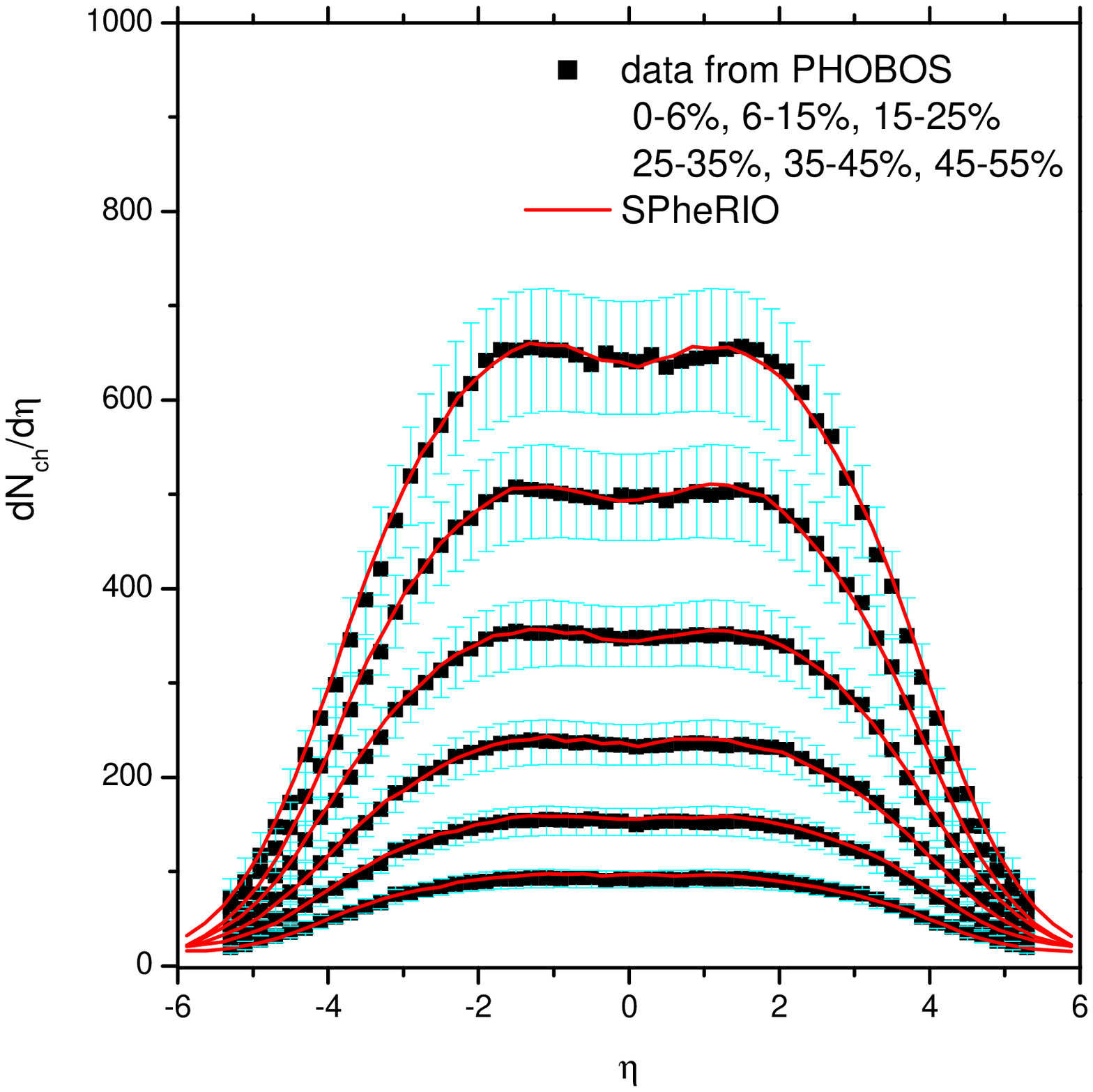}
\vspace*{-4cm}
\label{figure:fig1}
\caption{Pseudo-rapidity distributions of all charged particles for 
Au+Au collisions at 200A GeV. The data are from PHOBOS Collaboration\cite{phobos1}.} 
\end{figure}

\begin{figure}[!htb]
\vspace*{-1cm}
\includegraphics[width=8.5cm]{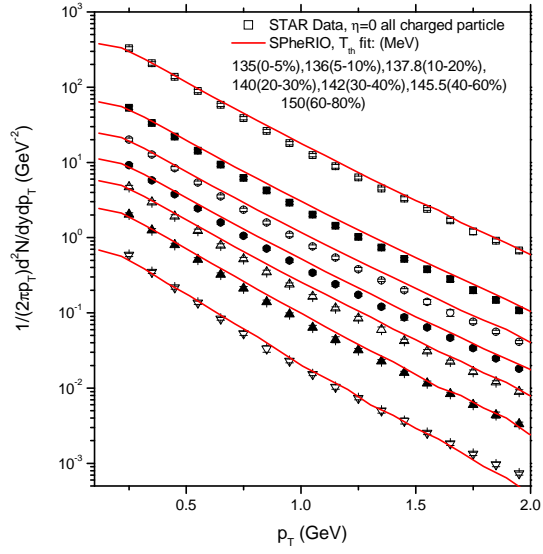}
\vspace*{-4cm}
\label{figure:fig2}
\caption{Transverse-momentum distributions of all charged particles for 
Au+Au collisions at 200A GeV in the pseudo-rapidity interval $-1.0 < \eta <
1.0$. The data are from STAR Collaboration\cite{star1}.} 
\end{figure}

\begin{figure}[!htb]
\vspace*{-1cm}
\includegraphics[width=8.5cm]{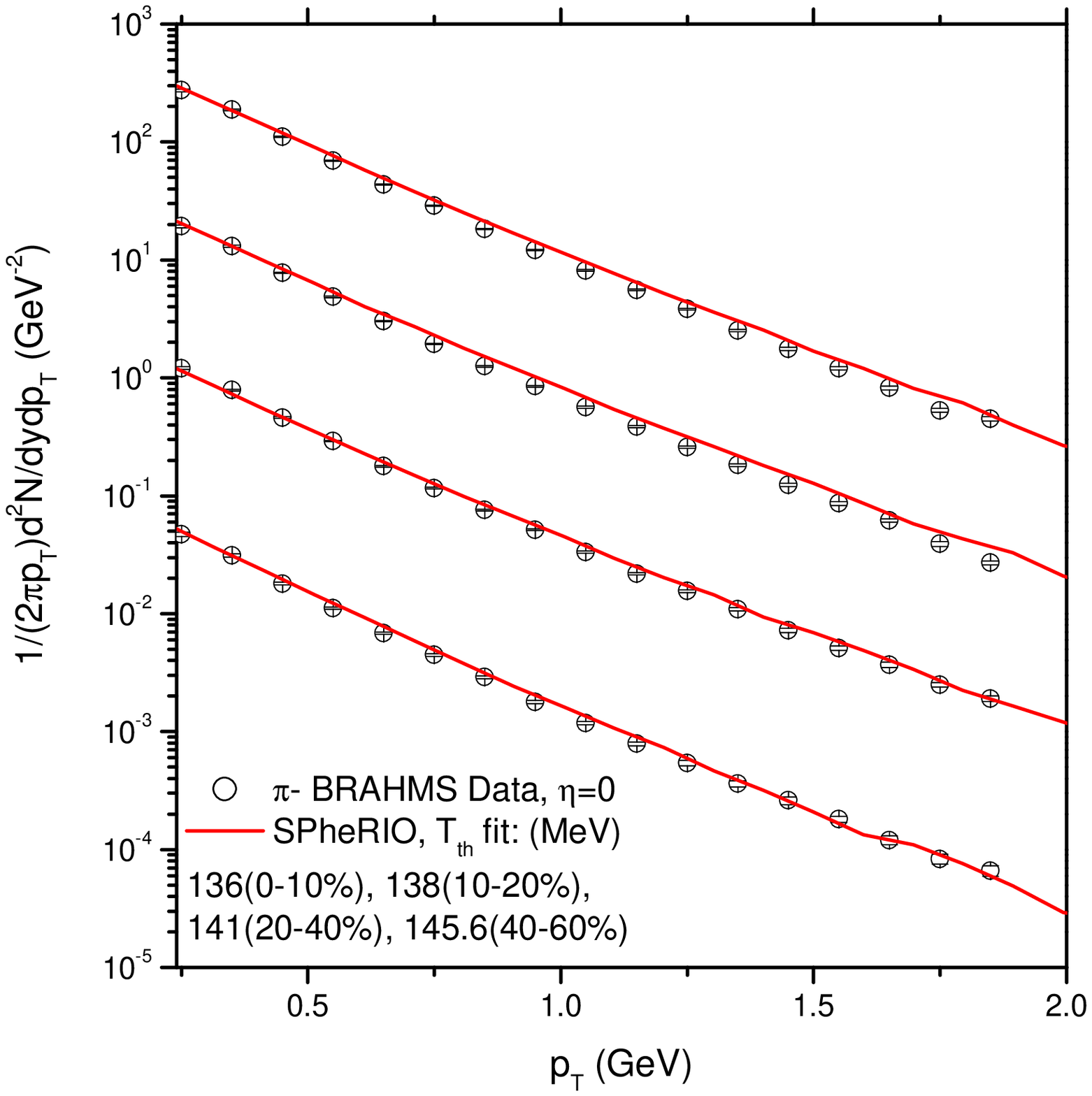}
\vspace*{-4cm}
\label{figure:fig3}
\caption{Transverse-momentum distributions of ${\pi}^-$ for 
Au+Au collision at 200A GeV in the pseudo-rapidity interval $-1.0 < \eta <1.0$. 
The data are from BRAHMS Collaboration\cite{brahms1}.} 
\end{figure}

\begin{figure}[!htb]
\vspace*{-1cm}
\includegraphics[width=8.5cm]{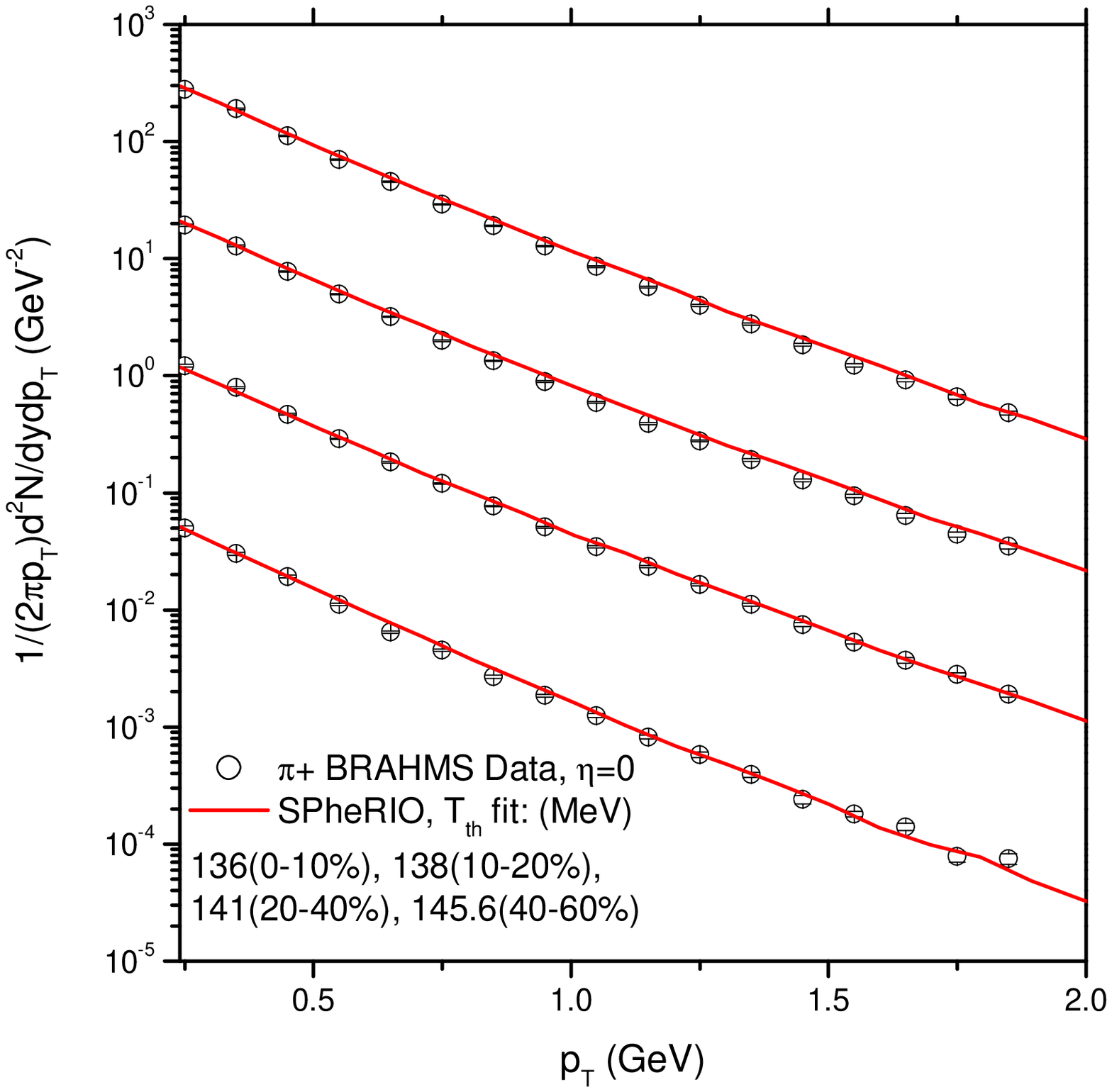}
\vspace*{-4cm}
\label{figure:fig4}
\caption{Transverse-momentum distributions of ${\pi}^+$ for 
Au+Au collisions at 200A GeV in the pseudo-rapidity interval $-1.0 < \eta <1.0$. 
The data are from BRAHMS Collaboration\cite{brahms1}.} 
\end{figure}

\begin{figure}[!htb]
\vspace*{-1cm}
\includegraphics[width=8.5cm]{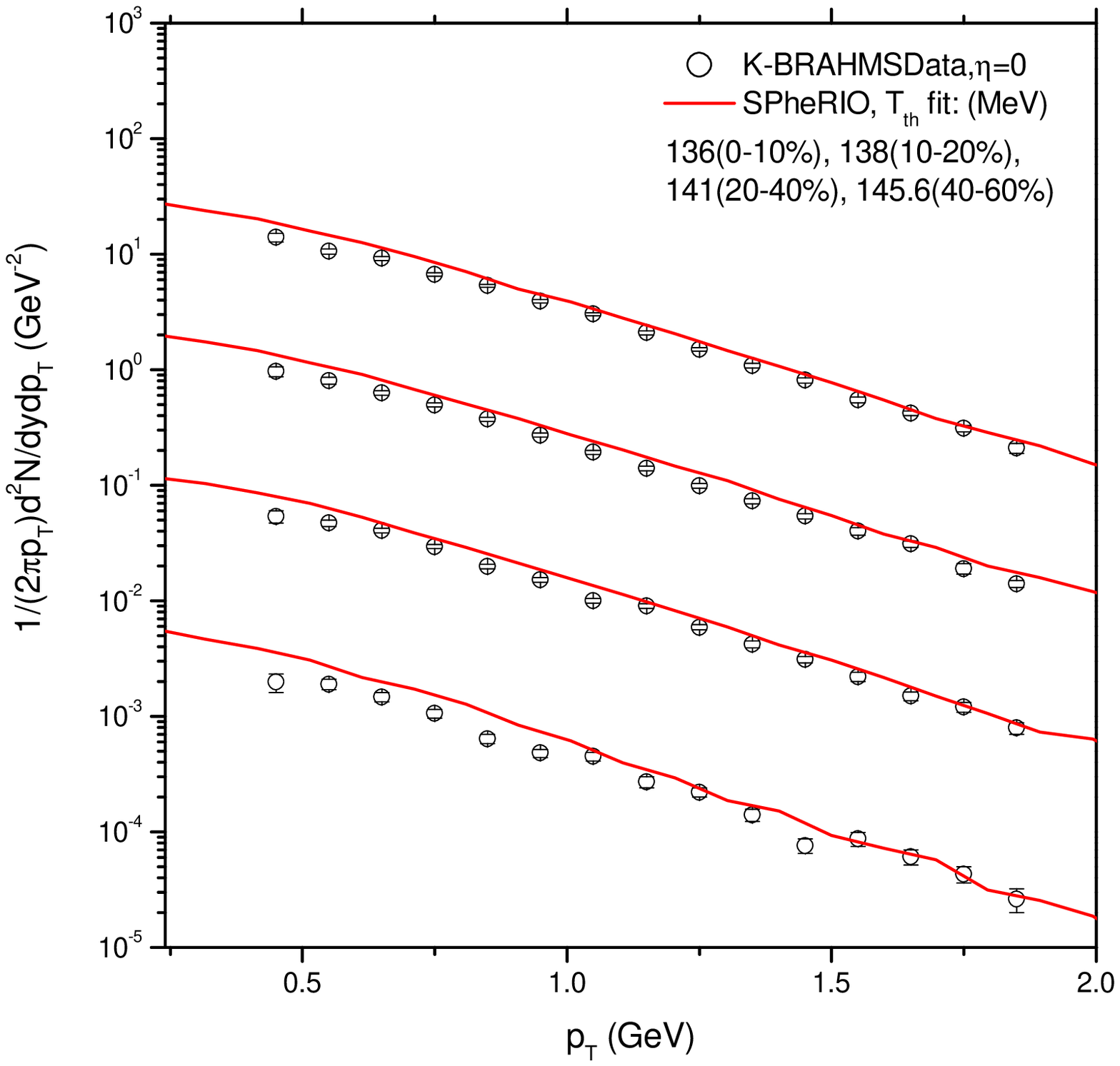}
\vspace*{-4cm}
\label{figure:fig5}
\caption{Transverse-momentum distributions of ${K}^-$ for 
Au+Au collisions at 200A GeV in the pseudo-rapidity interval $-1.0 < \eta <1.0$. 
The data are from BRAHMS Collaboration\cite{brahms1}.} 
\end{figure}

\begin{figure}[!htb]
\vspace*{-1cm}
\includegraphics[width=8.5cm]{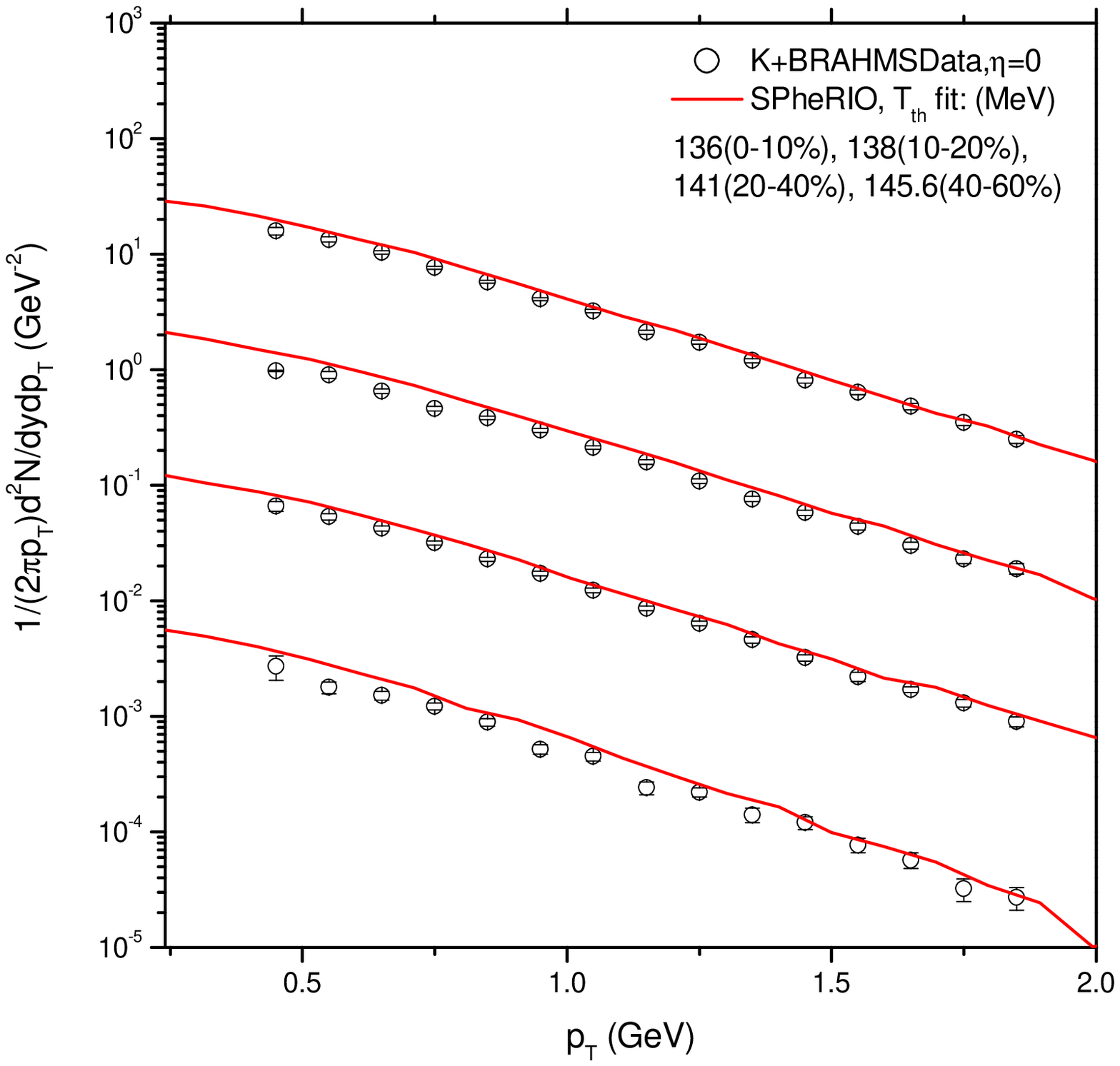}
\vspace*{-4cm}
\label{figure:fig6}
\caption{Transverse-momentum distributions of ${K}^+$ for 
Au+Au collisions at 200A GeV in the pseudo-rapidity interval $-1.0 < \eta <1.0$. 
The data are from BRAHMS Collaboration\cite{brahms1}.} 
\end{figure}

\begin{figure}[!htb]
\vspace*{-1cm}
\includegraphics[width=8.5cm]{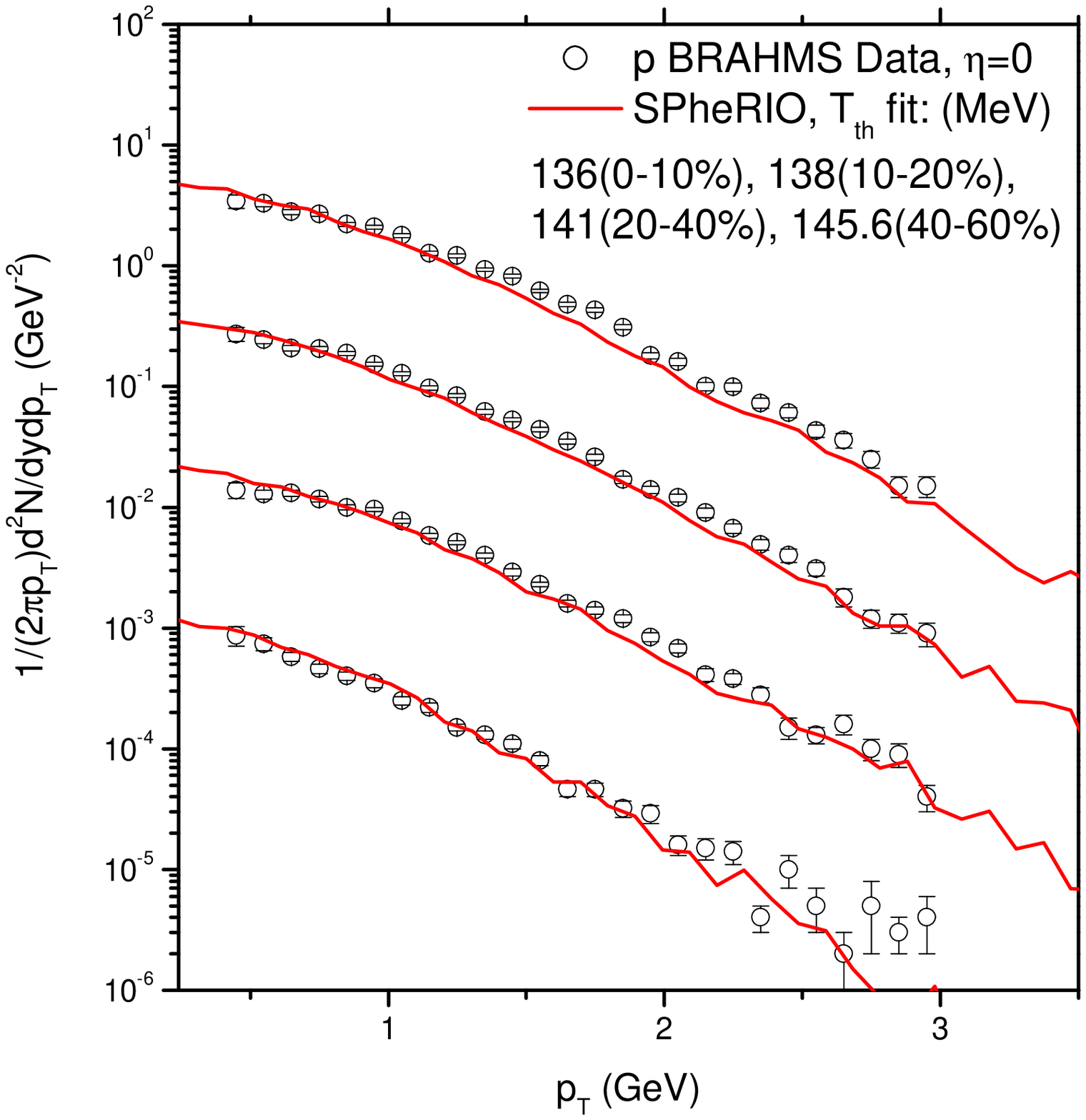}
\vspace*{-4cm}
\label{figure:fig7}
\caption{Transverse-momentum distributions of protons for 
Au+Au collisions at 200A GeV in the pseudo-rapidity interval $-1.0 < \eta <1.0$. 
The data are from BRAHMS Collaboration\cite{brahms1}.} 
\end{figure}

\begin{figure}[!htb]
\vspace*{-1cm}
\includegraphics[width=8.5cm]{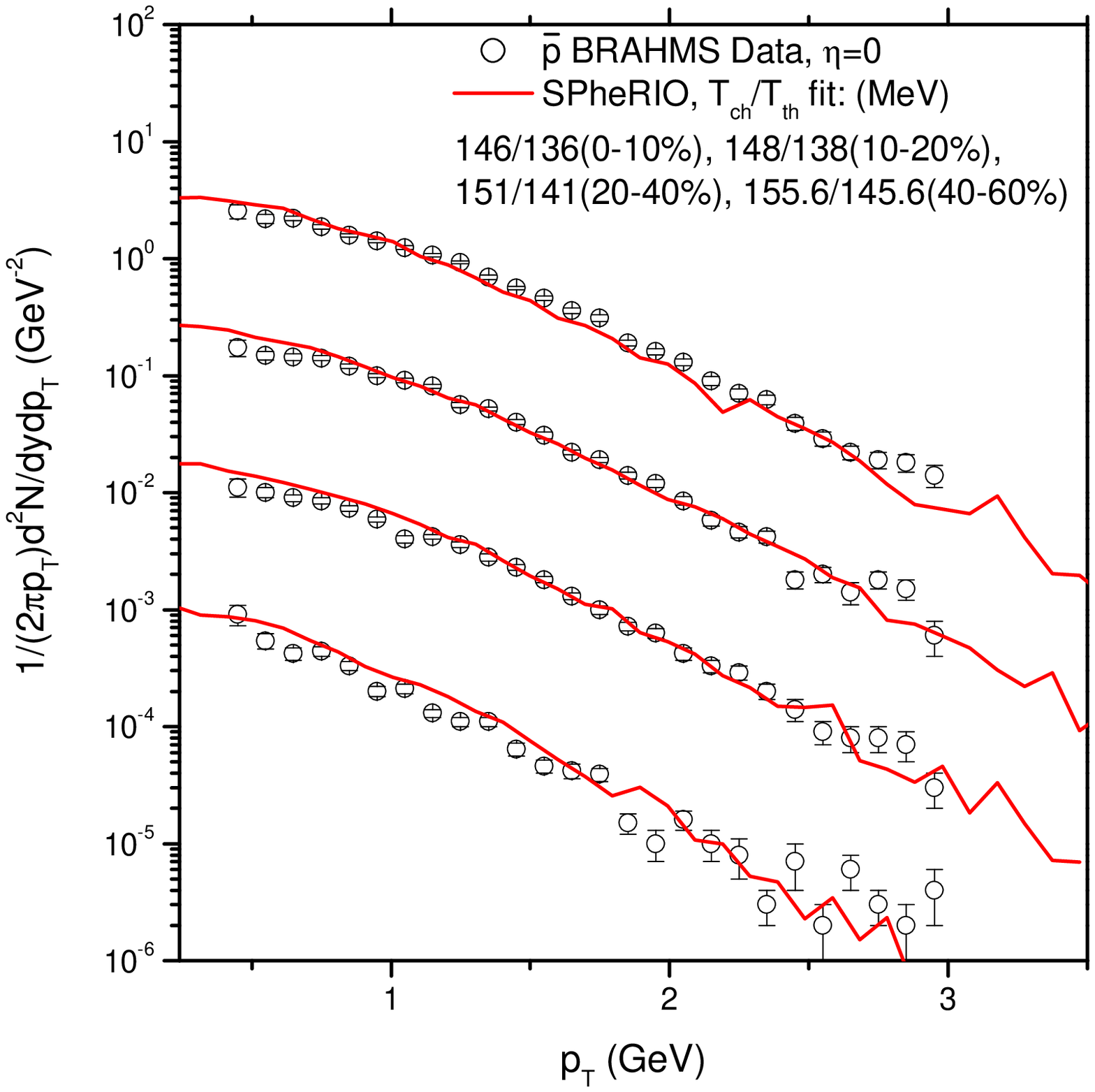}
\vspace*{-4cm}
\label{figure:fig8}
\caption{Transverse-momentum distributions of anti-protons for 
Au+Au collisions at 200A GeV in the pseudo-rapidity interval $-1.0 < \eta <1.0$. 
The data are from BRAHMS Collaboration\cite{brahms1}.} 
\end{figure}

\begin{figure}[!htb]
\vspace*{-1cm}
\includegraphics[width=8.5cm]{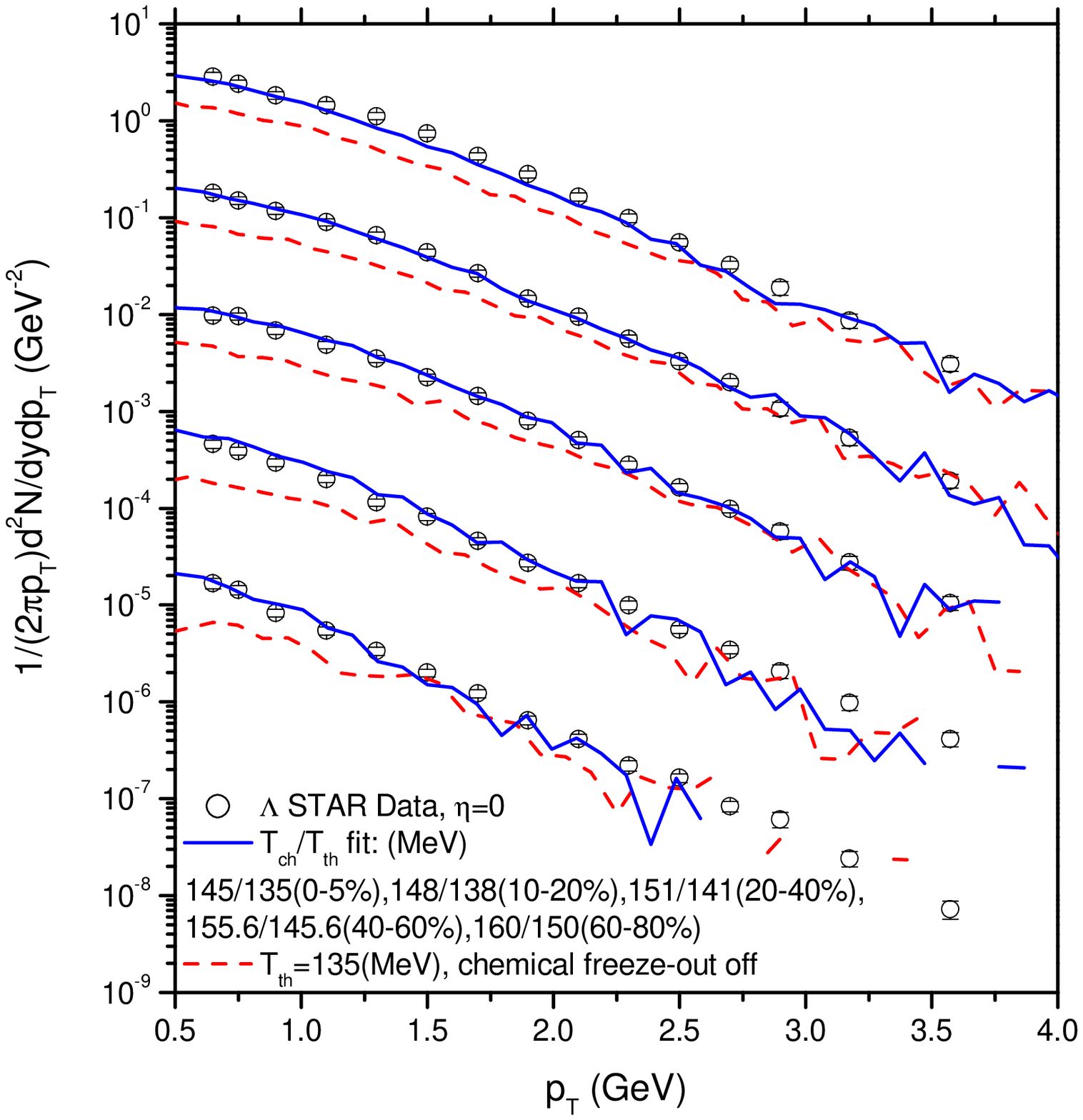}
\vspace*{-4cm}
\label{figure:fig9}
\caption{Transverse-momentum distributions of ${\Lambda}$ for 
Au+Au collisions at 200A GeV in the pseudo-rapidity interval $-1.0 < \eta <
1.0$. The data are from STAR Collaboration\cite{star2}. The dotted lines indicate
the results obtained without chemical freeze-out, and solid lines for
those with chemical freeze-out incorporated.} 
\end{figure}

\begin{figure}[!htb]
\vspace*{-1cm}
\includegraphics[width=8.5cm]{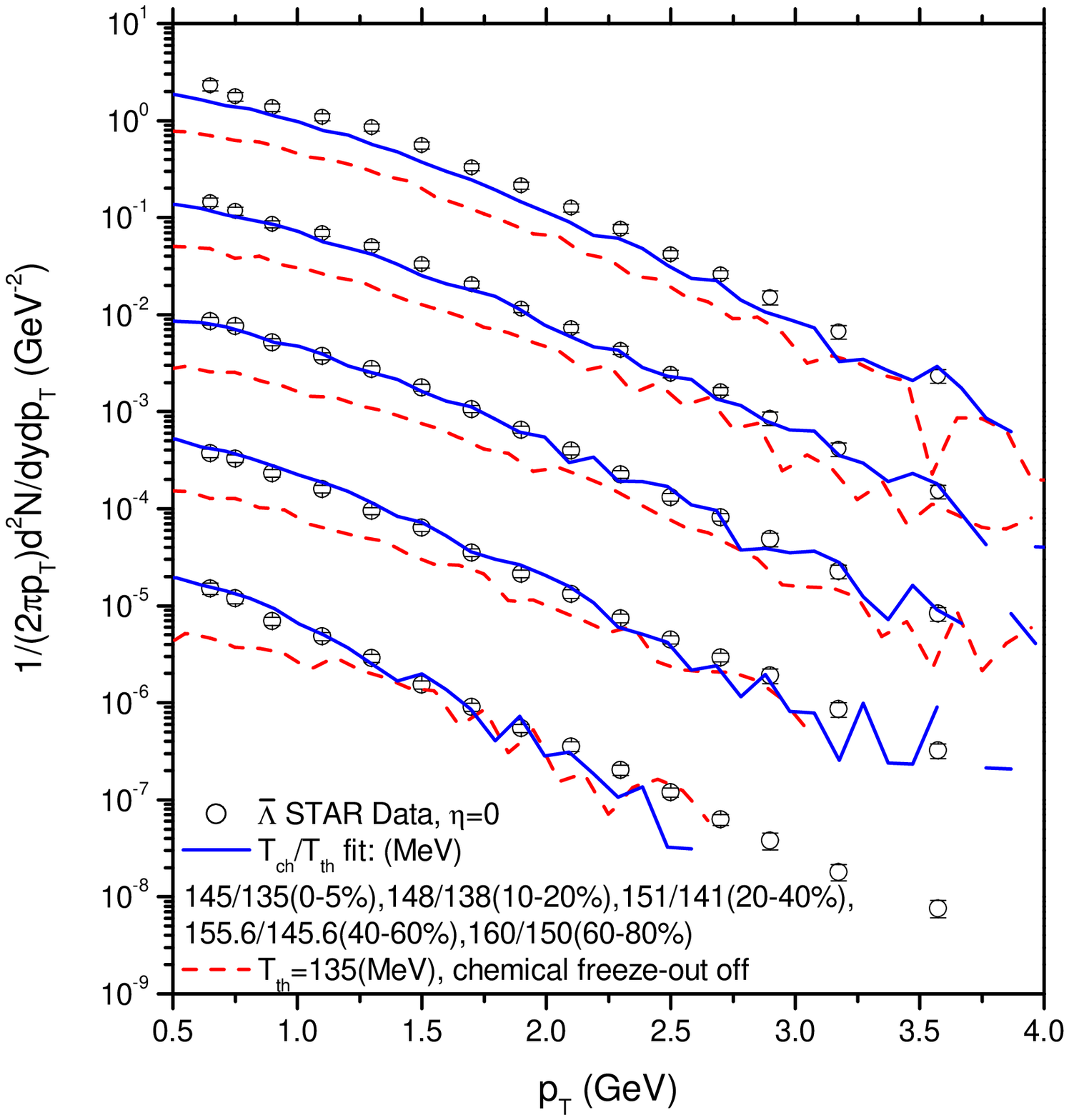}
\vspace*{-4cm}
\label{figure:fig10}
\caption{Transverse-momentum distributions of $\bar{\Lambda}$ for 
Au+Au collisions at 200A GeV in the pseudo-rapidity interval $-1.0 < \eta <
1.0$. The data are from STAR Collaboration\cite{star2}. The dotted lines indicate
the results obtained without chemical freeze-out, and solid lines for
those with chemical freeze-out incorporated.} 
\end{figure}

\begin{figure}[!htb]
\vspace*{-1cm}
\includegraphics[width=8.5cm]{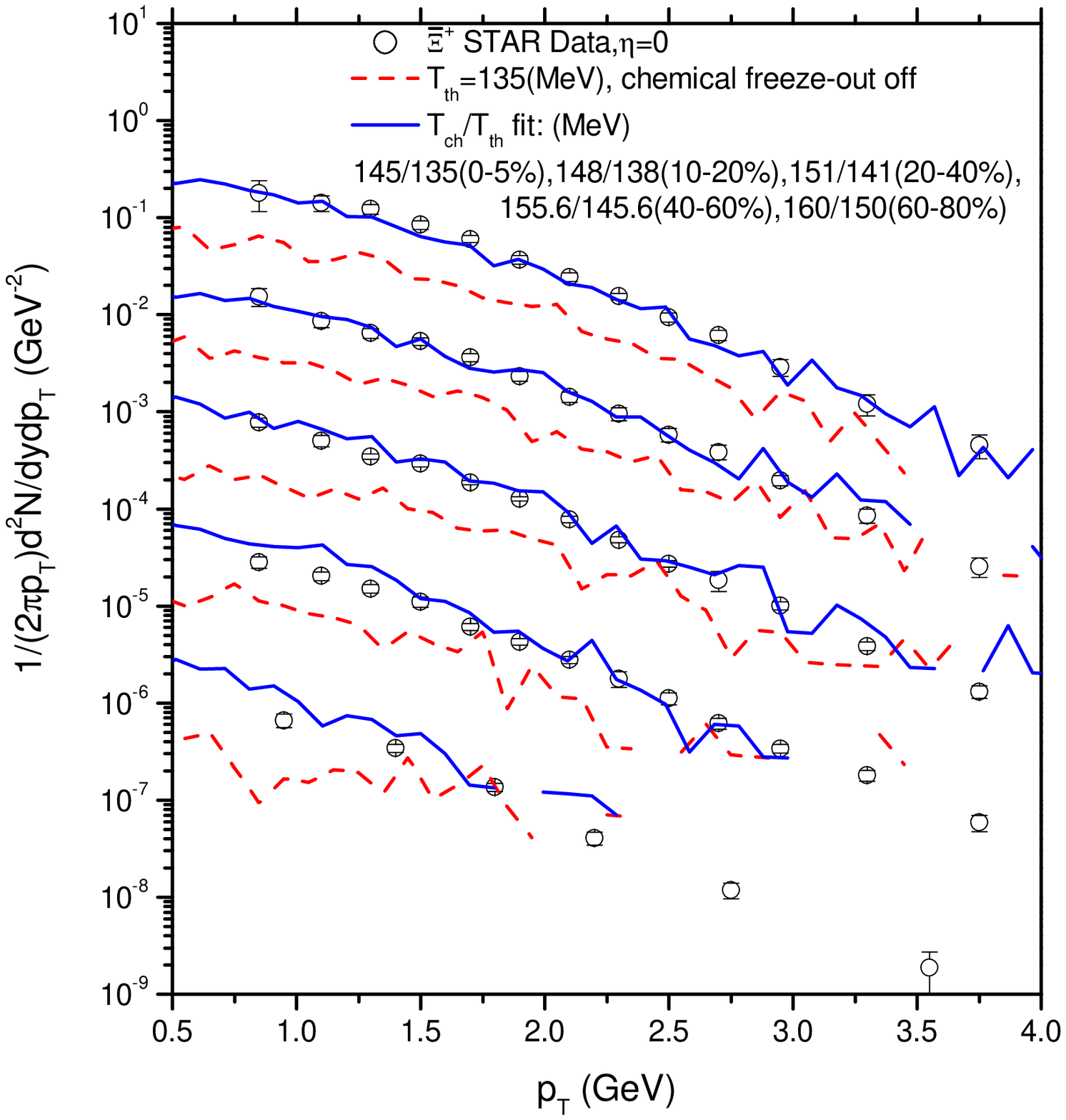}
\vspace*{-4cm}
\label{figure:fig11}
\caption{Transverse-momentum distributions of $\bar{\Xi}^+$ for 
Au+Au collisions at 200A GeV in the pseudo-rapidity interval $-1.0 < \eta <
1.0$. The data are from STAR Collaboration\cite{star2}. The dotted lines indicate
the results obtained without chemical freeze-out, and solid lines for
those with chemical freeze-out incorporated.} 
\end{figure}

\begin{figure}[!htb]
\vspace*{-1cm}
\includegraphics[width=8.5cm]{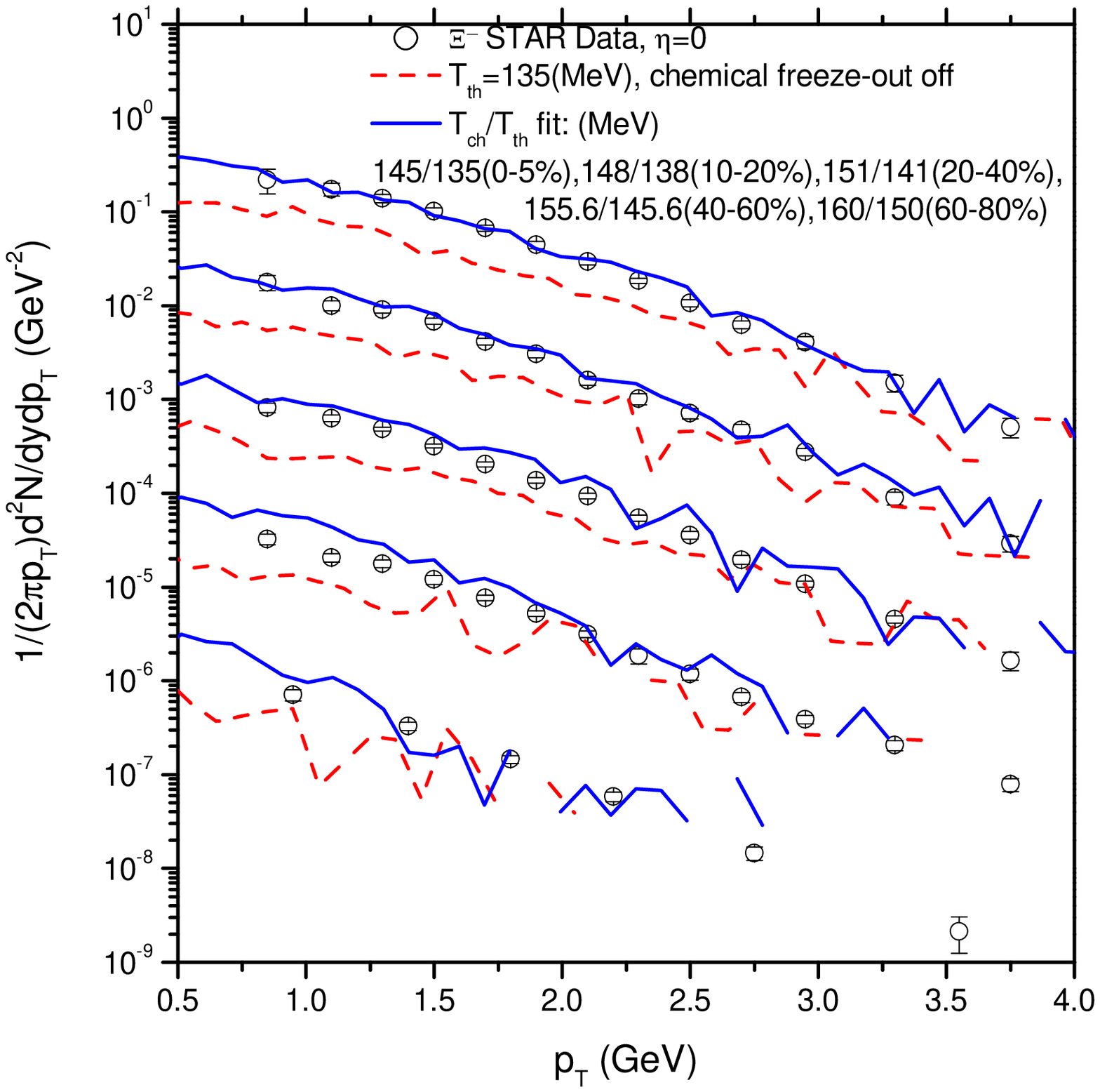}
\vspace*{-4cm}
\label{figure:fig12}
\caption{Transverse-momentum distributions of ${\Xi}^-$ for 
Au+Au collisions at 200A GeV in the pseudo-rapidity interval $-1.0 < \eta <
1.0$. The data are from STAR Collaboration\cite{star2}. The dotted lines indicate
the results obtained without chemical freeze-out, and solid lines for
those with chemical freeze-out incorporated.} 
\end{figure}

\begin{figure}[!htb]
\vspace*{-1cm}
\includegraphics[width=8.5cm]{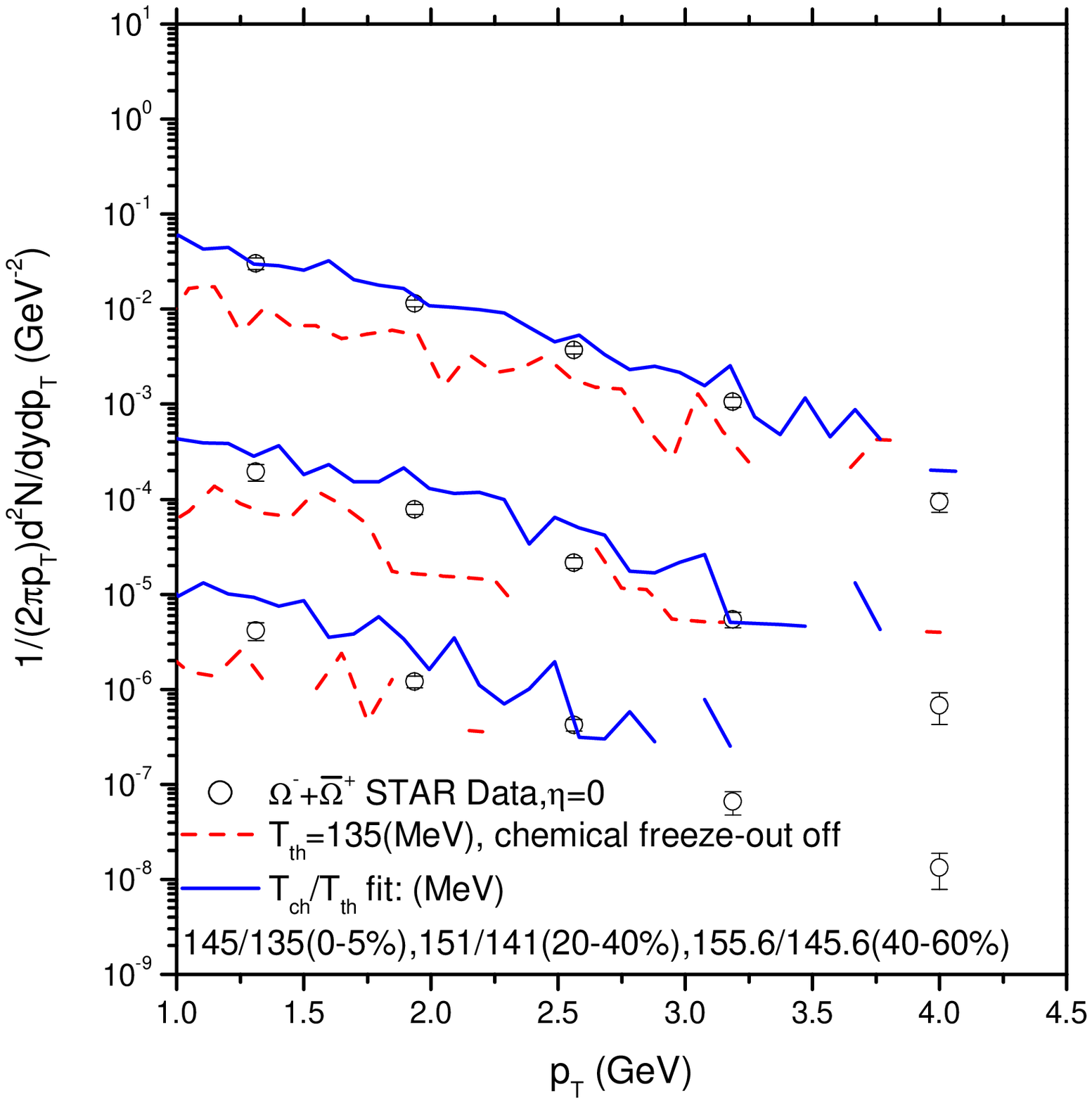}
\vspace*{-4cm}
\label{figure:fig13}
\caption{Transverse-momentum distributions of ${\Omega}^{-}+\bar{\Omega}^{+}$ for 
Au+Au collisions at 200A GeV in the pseudo-rapidity interval $-1.0 < \eta <
1.0$. The data are from STAR Collaboration\cite{star2}. The dotted lines indicate
the results obtained without chemical freeze-out, and solid lines for
those with chemical freeze-out incorporated.} 
\end{figure}

\end{document}